\documentclass[a4paper,twocolumn,showpacs,amsmath,pra,aps,10pt,amssymb]{revtex4-1} 
\usepackage[T1]{fontenc}
\usepackage[latin9]{inputenc} 
\usepackage{times}
\usepackage{color}
\usepackage{bm}
\usepackage{xspace}
\usepackage{amssymb,amsmath}
\usepackage{amsbsy}
\usepackage[pdftex]{graphicx}
\usepackage{float}

\newcommand{\br}{\mathbf{r}}
\newcommand{\bx}{\mathbf{x}}

\newcommand{\UD}{U_{\mathrm{dd}}}

\begin{document}
\title{Roton spectroscopy in a harmonically trapped dipolar Bose-Einstein condensate} 
\author{P.~B.~Blakie}  
\author{D.~Baillie}  
\author{R.~N.~Bisset}  

\affiliation{Jack Dodd Centre for Quantum Technology, Department of Physics, University of Otago, Dunedin, New Zealand.}

\begin{abstract} 
We study a harmonically trapped Bose-Einstein condensate with dipole-dipole interactions in a regime where a roton spectrum emerges.  We show that the roton spectrum is clearly revealed in the static and dynamic structure factors which can be measured using Bragg spectroscopy. We develop and validate a theory based on the local density approximation  for the dynamic structure factor.   
\end{abstract}


\maketitle

 
A variety of atomic species with large magnetic dipoles have been cooled to the quantum degenerate regime in which a  Bose-Einstein condensate (BEC)  forms \cite{Griesmaier2005a,*Bismut2010a,*Mingwu2011a,*Aikawa2012a}. The key feature of these gases is that the particles interact via a dipole-dipole interaction (DDI) that is long-ranged and anisotropic \cite{Lahaye2007a}.   
A fascinating prediction is that when such a BEC is tightly confined in the direction that the dipoles are polarized, a roton-like spectrum can emerge \cite{Santos2003a,*ODell2003a,*Fischer2006a,*Ronen2007a}. A number of theoretical proposals for detecting roton features have been made including sensitivity to external perturbations \cite{Wilson2009a}, depression in the critical velocity \cite{Wilson2011b,Ticknor2011a}, and signatures in density fluctuations \cite{Klawunn2011a} (c.f.~\cite{Hung2011b}).      However, to date there has been no experimental evidence for roton properties in dipolar BECs  \footnote{Roton mode softening has been observed in a BEC with cavity mediated long-range interactions \cite{Mottl2012a}.}.

In this paper we study a dipolar BEC confined in a quasi-two-dimensional (quasi-2D) harmonic trap. We vary contact and dipole interaction parameters over a wide range and characterize the emergence of a roton through the static and dynamic structure factors. These quantities closely relate to the observable for Bragg spectroscopy \cite{Stenger1999a,*Stamper-Kurn1999a,*Steinhauer2002a,Zambelli2000a}, and thus are readily measured in experiments. We note that Bragg spectroscopy has emerged as a flexible tool for investigating ultra-cold gases and has been applied to resonant Bose \cite{Papp2008a} and Fermi \cite{Veeravalli2008a} gases, quasi-1D Bose gases \cite{Richard2003a}, and vortices in BECs \cite{Blakie2001a,*Muniz2006a}. Recently the first application of Bragg spectroscopy to a dipolar  BEC has been made \cite{Bismut2012a} in a nearly spherical trap, and used to demonstrate an anisotropic speed of sound.

Our calculations for the structure factors are based on solving the non-local Gross-Pitaevskii equation (GPE) for the condensate and the Bogoliubov de-Gennes (BdG) equations for the quasi-particle excitations. Our calculations are fully three-dimensional (3D), i.e.~we do not make the quasi-2D approximation in which an ansatz for the condensate shape in the tightly confined direction is assumed (this approximation has been shown to be surprisingly inaccurate in the regime of interest \cite{Wilson2011a}).  
We finish by developing a local density approximation (LDA)  theory that provides a reasonably accurate description of our full theory. We emphasize that the regime of our study is appropriate to current experiments with magnetic dipoles. {We refer the reader to Ref.~\cite{Filinov2012a} for a discussion of  the pure-2D regime that might be realized in future polar-molecule experiments.}

 The DDI potential between a  pair of dipolar atoms is
\begin{align}
\UD(\br)=\frac{3g_{\mathrm{dd}}}{4\pi}\frac{1-3\cos^2\theta}{|\br|^3},\label{e:udd}
\end{align}
where $g_{\mathrm{dd}}=\mu_0\mu^2_m/3$,  $\mu_m$ is the dipole strength, and $\theta$ is the angle between the dipole separation $\br$ and the polarization axis, which we take to be the $z$ direction. The atoms also interact via a contact interaction of strength $g=4\pi a\hbar^2/m$ with $a$ the $s$-wave scattering length.  We take the  atoms to be confined in a cylindrically symmetric trap $U(\bx)=\tfrac{1}{2}m\omega_{\rho}^2(\rho^2+\lambda^2z^2)$  of aspect ratio $\lambda=\omega_z/\omega_\rho$. Here we consider tight axial confinement  ($\omega_z\gg\omega_{\rho}$) to produce a quasi-2D trap favorable   for the emergence of rotons.

\begin{figure}[htbp] 
   \centering
   \includegraphics[width=3.4in]{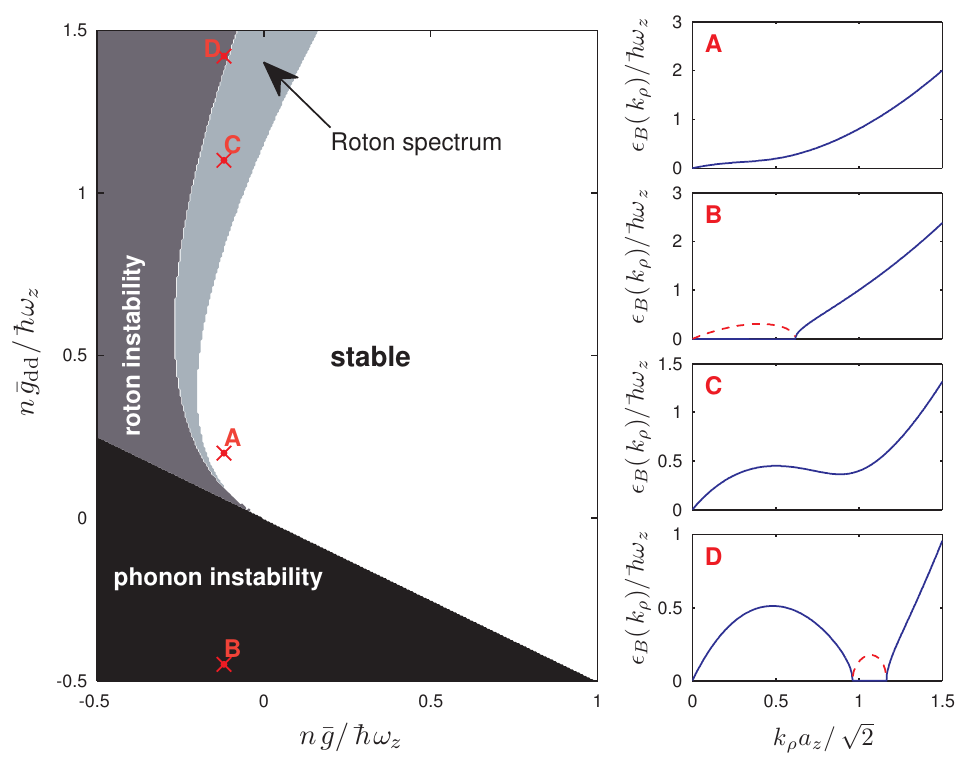}  
   \caption{(color online) Stability phase diagram and related excitation properties of quasi-2D uniform dipolar BEC. White and light-grey  regions indicate where the BEC is dynamically stable. In the light-grey region  the spectrum has a  roton minimum. In the dark-grey and black regions the system is dynamically unstable. This can arise from modes at zero  momentum (phonon instability -- black region) or finite momentum (roton instability -- dark grey region) developing imaginary parts. Subplots A-D show cases of the  spectrum (\ref{Eubog}), with the real  (solid line) and imaginary (dashed line) parts shown.  
   }
   \label{fig:2D}
\end{figure}
It is useful to  review the properties of a homogeneous dipolar gas (i.e.~with $\omega_{\rho}=0$), which admits a simple analytic treatment under the quasi-2D approximation (assuming a Gaussian mode structure along the tight direction). In this case the tight direction can be integrated out and  the Fourier transform of the  in-plane interaction potential is \cite{Fischer2006a}
\begin{equation} 
\tilde{V}_{\mathrm{2D}}({{k}_\rho})=\bar{g}+ \bar{g}_{\mathrm{dd}} F_{\perp}(k_\rho a_z/\sqrt{2}),\label{V2D}
\end{equation}
where $
F_{\perp}(x)=2-3\sqrt{\pi}xe^{x^2}\mathrm{erfc}\left(x\right)$,  $\bar{g}=g/\sqrt{2\pi}a_z$ and  $\bar{g}_{\mathrm{dd}}=g_{\mathrm{dd}}/\sqrt{2\pi}a_z$ are the 2D interaction parameters, with $a_z=\sqrt{\hbar/m\omega_z}$ the $z$-confinement length scale. For a condensate of areal density $n$ the Bogoliubov dispersion relation for in-plane modes  (i.e.~no $z$ excitation) is  
\begin{equation}
\epsilon_{\mathrm{B}}( {k}_\rho)=\sqrt{\epsilon(k_{\rho})^2 +2\epsilon(k_{\rho}) {n}\tilde{V}_{\mathrm{2D}}({{k}_\rho}) },\label{Eubog}
\end{equation} 
where   $\epsilon(k_{\rho})=\hbar^2 k^2_{\rho}/{2m}$. 
In Fig.~\ref{fig:2D} we show the generic features of the quasi-2D system as the dipolar and contact interaction parameters are varied.
Notably, the system can become unstable through a phonon   or roton  instability where $k_{\rho}\to0$ (case B) or $k_{\rho}\sim1/a_z$ (case D) modes, respectively, soften and develop imaginary eigenvalues. Within the stable region we have indicated a sub-region where the dispersion relation has a roton feature i.e.~a local minimum at finite $k_{\rho}$ (case C).

We now turn to our main concern, a fully 3D calculation of the trapped system. To do this we numerically solve the non-local GPE for the unit normalized condensate orbital $\psi_0(\mathbf{x})$ and chemical potential $\mu$, and then diagonalize the BdG equations for the  quasi-particle excitations $\{u_j(\mathbf{x}),v_j(\mathbf{x})\}$, with respective energies $\epsilon_j$ \footnote{We follow the notation and conventions for the dipolar GPE and BdG equations given in Ref.~\cite{Ronen2006a}.}. Our numerical method, similar to Ref.~\cite{Ronen2006a}, utilizes the cylindrical symmetry by employing a Fourier-Hankel representation to separate the GPE and BdG equations into a set of 2D problems specified by the angular quantum number $m_z=0,\pm1,\pm2,\ldots$ [e.g.~$u_j(\mathbf{x})=u_j(\rho,z)e^{im_z\phi}$]. We also use a cylindrically  cutoff of the interaction potential (\ref{e:udd})   to improve numerical convergence \cite{Lu2010a}.

The results we present focus on  a trap with $\lambda=40$, although we find qualitatively similar behavior   for   $\lambda\gtrsim10$. We have chosen $\lambda=40$ as being sufficiently tight  for the roton to emerge at a reasonably large $k$ value, yet is an aspect ratio that  is readily achievable in experiments (e.g.~\cite{Clade2009a}).
For convenience we introduce $D=3Ng_{\mathrm{dd}}m/4\pi \hbar^2a_{\rho}$ and $C=Ngm/4\pi\hbar^2a_{\rho}$ as dimensionless  parameters for the dipolar and contact interactions, respectively, with $a_{\rho}=\sqrt{\hbar/m\omega_{\rho}}$ and $N$ the number of condensate atoms. 

\begin{figure}[htbp] 
   \centering 
   \includegraphics[width=3.3in]{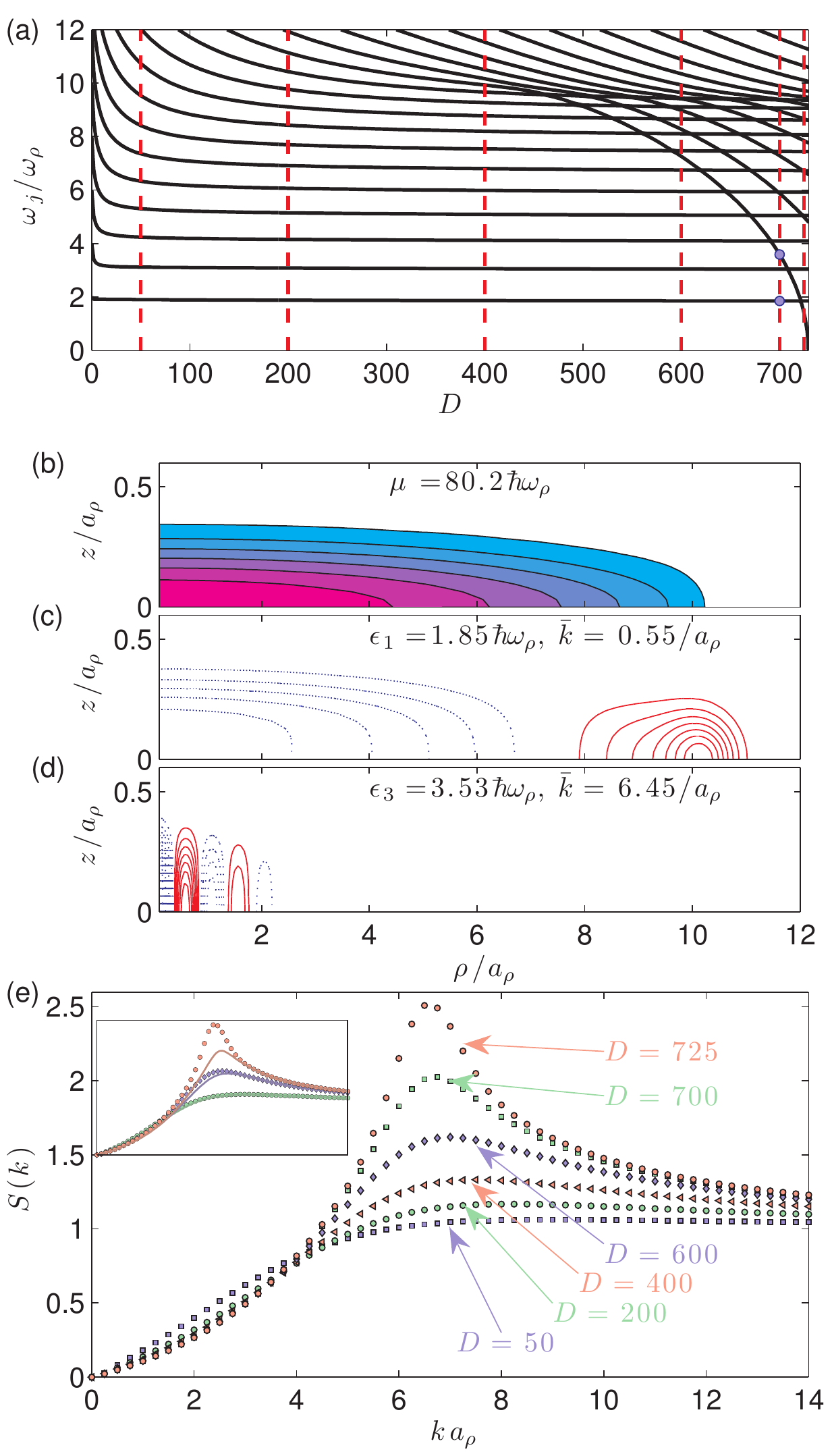} 
    \vspace*{-0.5cm}\caption{(color online) (a) Spectrum of $m_z=0$ quasi-particle modes  demonstrating the roton-softening.   (b)-(d) Contour plots of modes  at $D=700$. (b) Condensate density. (c) and (d) give  the density perturbation [$\delta \rho_j\!=\!\psi_0(u_j\!+\!v_j)$]   for  the first [(c)] and  third (roton) [(d)]   $m_z=0$ quasi-particles [these modes are indicated by circles   in (a)]. Solid  (dotted) contours indicate positive (negative)  density  perturbations. 
    (e)    $S(k)$ for  $D$ values corresponding to the vertical dashed lines in (a).  Inset: A comparison of full   calculations for $S(k)$ (symbols) to the LDA results  [Eq.~(\ref{Sqw_GPELDA})] (lines) for    $D=200,\,600,\,725$ (lowest to higher curves). Results for  $\lambda=40$ and $C=0$. }
   \label{fig:Sqtrapped}
\end{figure}

In Fig.~\ref{fig:Sqtrapped}(a) we show the excitation spectrum of $m_z=0$ modes  for $C=0$ as a function of $D$. For $D\gtrsim400$ we observe that high-lying quasi-particle modes begin to rapidly decrease in energy as $D$   increases.  We identify the softening of these highly excited modes (i.e.~modes with many radial nodes) as the manifestation of the roton spectrum in the trapped gas. This interpretation is supported by other studies of the trapped system in which the quasi-particle spectrum was approximately mapped onto  a dispersion relation \cite{Wilson2010a} (also see $\bar{k}$ defined below).  At $D\approx730$ the first of these {roton} (quasi-particle) modes hits zero energy and develops an imaginary part, signaling the onset of a dynamical instability. We note that modes with $|m_z|>0$   exhibit similar trends to the $m_z=0$ results shown in Fig.~\ref{fig:Sqtrapped}(a). 
 
The condensate orbital and the density perturbations associated with two quasi-particle modes are indicated in Fig.~\ref{fig:Sqtrapped}(b)-(d). Notably the roton mode [Fig.~\ref{fig:Sqtrapped}(d)] is localized near the center of the condensate and has a short wavelength. Following \cite{Wilson2010a} we assign a wavevector to the modes according to $\bar{k}=\sqrt{\langle k_{\rho}^2\rangle}$, and find that for this mode $\bar{k}=6.45/a_{\rho}$, similar to the inverse $z$ confinement length   $1/a_z\approx6.3/a_\rho$.

The dynamic structure factor for a $T=0$ BEC is \cite{Zambelli2000a},
\begin{equation}
S(\mathbf{k},\omega)\!=\!\sum_j\!\Big|\int \!d\bx\,[u_j^*(\bx)+v_j^*(\bx)]e^{i\mathbf{k}\cdot\bx}\psi_0(\bx)\Big|^2 \!\delta(\omega\!-\!\omega_j),\label{Skw}
\end{equation}
where   $\omega_j\equiv\epsilon_j/\hbar$. 
It is worth noting that Bragg spectroscopy measures the imaginary part of the response function  $\mathrm{Im}\left[\chi_{\mathbf{k}}(\omega)\right]=-\tfrac{\pi}{\hbar}[S(\mathbf{k},\omega)-S(-\mathbf{k},-\omega)]$ \cite{Brunello2001a,*Blakie2002a}, and to leading order this  is only sensitive to the zero-temperature dynamic structure factor. Thus our results should be  applicable to regimes with a discernible non-condensate fraction. Corrections beyond leading order will require a finite temperature extension of the theory (e.g.~see \cite{Ronen2007b,*Bisset2011a,*Ticknor2012a}).

Integrating $S(\mathbf{k},\omega)$ over frequency yields the static  structure factor 
$S(\mathbf{k})= \int d\omega\,S(\mathbf{k},\omega)$,
which also relates to the Fourier transform of the pair correlation function \cite{Zambelli2000a}.
For the uniform system $S(\mathbf{k})$ directly gives the dispersion relation  through the Bijl-Feynman formula 
$S(\mathbf{k})= {\epsilon(\mathbf{k})}/{\epsilon_B(\mathbf{k})}$. 

Here we restrict our attention to evaluating the structure factors for radial wavevectors $\mathbf{k}=k\,\hat{\mathbf{k}}_{\rho}$, since the roton modes exhibit non-trivial structure in-plane  [see Fig.~\ref{fig:Sqtrapped}(d)], and from hereon will denote these  with scalar arguments, i.e.~$S(k,\omega)$, $S(k)$.
For a given value of $k$ the numerical evaluation of $S(k,\omega)$  requires including modes up to a maximum energy $\epsilon_{\max}$ with $\epsilon_{\max}\gtrsim\hbar^2k^2/2m$. In practice we check that sufficiently many modes are included by ensuring that  the $f$-sum rule, $\int_0^\infty d\omega \,\omega S(\mathbf{k},\omega)= {\hbar k^2}/{2m}$, is satisfied. For the $k$ values we consider here ($k\lesssim20/a_{\rho}$) we typically use $\gtrsim10^4$ modes in our calculations. 

We present results for $S(k)$  in Fig.~\ref{fig:Sqtrapped}(e) for various values of the dipole interaction strength.
The suppression of $S(k)$ as $k\to0$ reveals the low-energy phonon spectrum of the system (see \cite{Stamper-Kurn1999a}). A  significant peak in $S(k)$ at $k_{\mathrm{peak}}\sim6.5/a_{\rho}$  forms for interaction values of  $D>400$, which corresponds to where the high energy modes begin to  rapidly descend in the spectrum [see Fig.~\ref{fig:Sqtrapped}(a)].    Appealing to the Bijl-Feynman formula we identify a significant peak in the static structure factor with the appearance of a roton feature in the excitation spectrum.
This identification is useful because it corresponds to a practical experimental observable and does not depend upon any  \textit{ad hoc} scheme for assigning a dispersion relation to the excitations of the trapped system.
 \begin{figure}[!htbp] 
   \centering
   \includegraphics[width=3.4in]{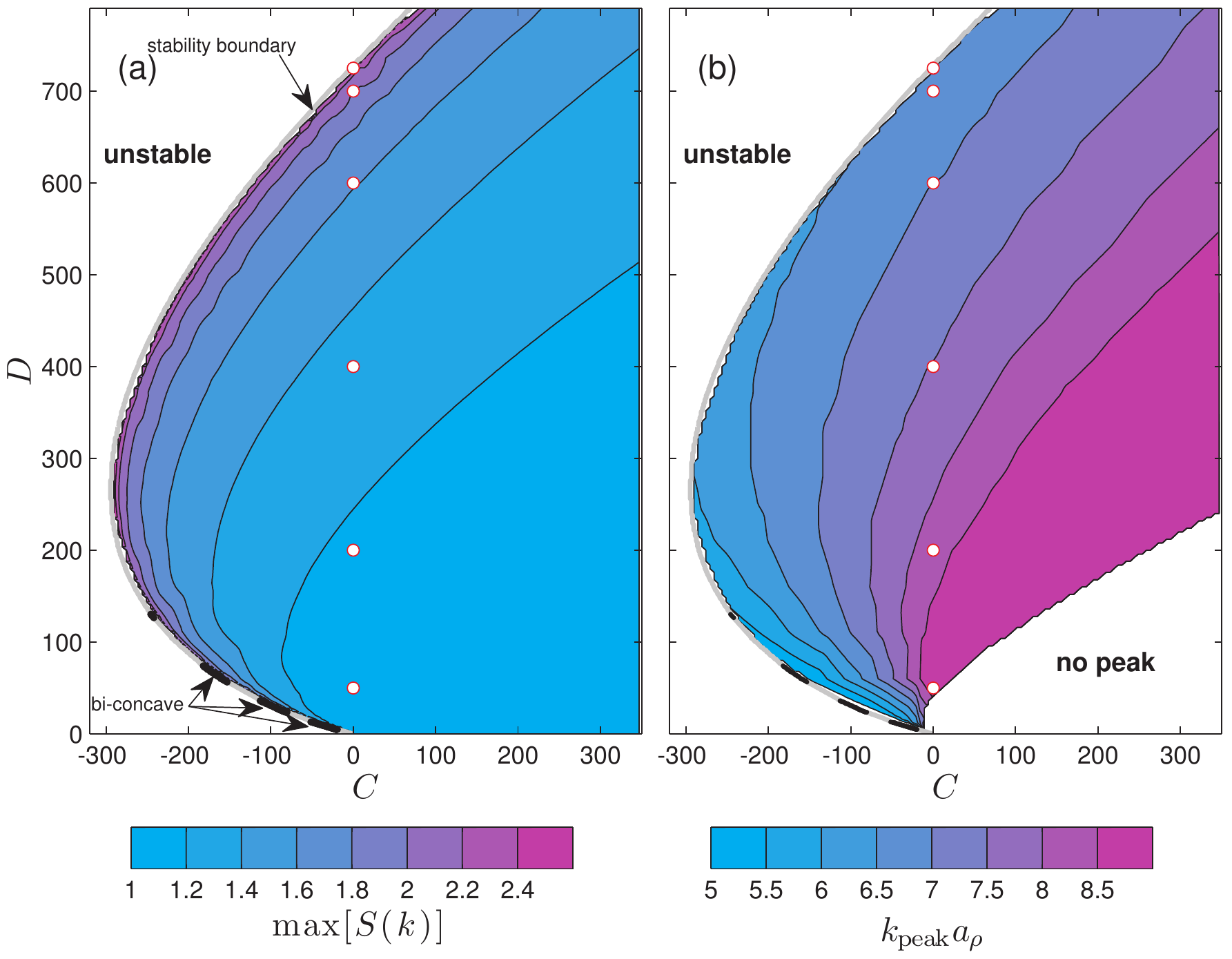}  
   \vspace*{-0.25cm}\caption{(color online) Characterization of the roton properties of a   $\lambda=40$ trapped dipolar BEC using  $S(k)$.
   (a)  Peak value of $S(k)$. (b) Peak wavevector $k_{\mathrm{peak}}$ (none shown when $\max[S(k)]<1.05$).
    Circles mark the parameters of states analyzed in Fig.~\ref{fig:Sqtrapped}(e). Grey boundary line indicates when the system is unstable due to excited modes softening, with black segments indicating that the BEC is in a bi-concave state at the boundary  (see \cite{Ronen2007a}). }
   \label{fig:contours}
\end{figure}

In Fig.~\ref{fig:contours}  we characterize the behavior of $S(k)$ over a broad range of contact and dipole parameters where the BEC is dynamically stable. We show where a peak in $S(k)$ emerges and characterize its height [Fig.~\ref{fig:contours}(a)] and wavevector ($k_{\mathrm{peak}}$) [Fig.~\ref{fig:contours}(b)]. 
Our results show that the roton character of the spectrum is generally enhanced [i.e.~height of peak in $S(k)$ increases] at fixed dipole strength by decreasing (i.e.~making more negative) the contact interaction strength, although the value of $k_{\mathrm{peak}}$ tends to decrease as this happens. 

In Fig.~\ref{fig:contours}  we also indicate the boundary upon which the BEC becomes dynamically unstable. Because the cloud is tightly confined in the $z$ direction the repulsive character of the DDI (due to side-by-side dipoles) dominates and hence the DDI partially stabilizes the BEC against collapse from a negative value  of the contact interaction. Similar observations, for gases with smaller trap aspect ratio, were presented in Ref.~\cite{Lu2010a}. 
  There are regions near the boundary where the condensate develops a bi-concave density profile with a local minimum in the BEC density  at trap centre \cite{Ronen2007a}\footnote{We only show where the bi-concave regions intercept with the stability boundary.}. We do notice  any  signature of the bi-concave BEC in $S(k)$.
  
\begin{figure}[!htbp] 
   \centering
   \includegraphics[width=3.4in]{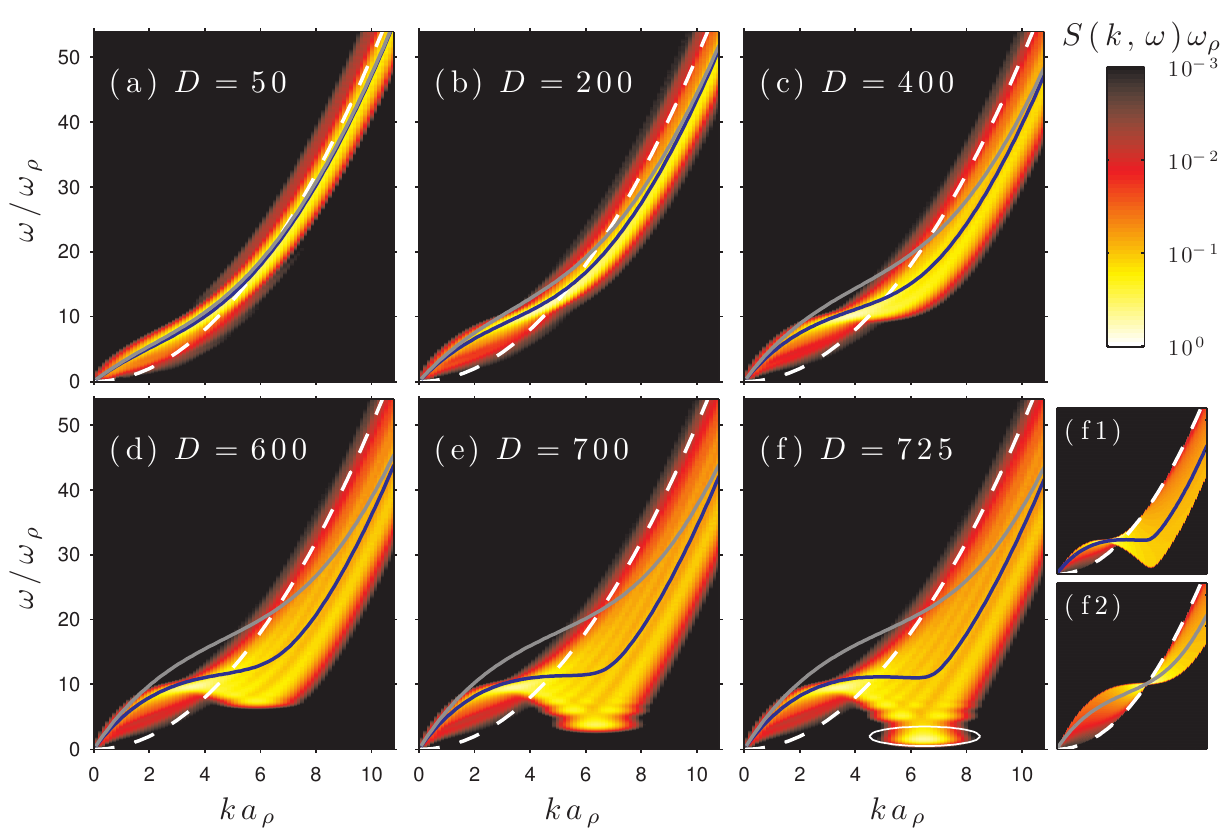} 
     \vspace*{-0.25cm} \caption{(color online) (a)-(f) $S(k,\omega)$ for indicated values of $D$. A  Gaussian of width $\Delta\omega=0.5\,\omega_\rho$ is used to smooth  the $\delta$-functions in $S(k,\omega)$. 
     The white dashed line shows the free particle dispersion $\epsilon(k)/\hbar$. The solid   black and grey lines are the mean response  $\bar{\omega}(k)$ obtained from GPE based and quasi-2D approximation LDA calculations, respectively  [see text and below]. White ellipse in (f) identifies a roton response feature.
 LDA calculations of $S(k,\omega)$ for the parameters  in (f) are shown in (f1) and (f2). (f1) GPE based LDA approach of  Eq.~(\ref{Sqw_GPELDA}).  (f2) quasi-2D approximation LDA approach (see text). The mean response  $\bar{\omega}(k)\equiv\int d\omega\,\omega S(k,\omega)/S(k)$,  is shown for each result. Other parameters:  $C=0$ and $\lambda=40$.}
   \label{fig:Sqwtrapped}
\end{figure}

Because the frequency dependence of the system response is most directly measured in Bragg spectroscopy experiments it is also worth discussing the behavior of the dynamic structure factor. 
In Figs.~\ref{fig:Sqwtrapped}(a)-(f) we show $S(k,\omega)$  for the same cases considered in Fig.~\ref{fig:Sqtrapped}(e). In the vicinity of  the roton wavevector [i.e.~$k\sim6.5/a_{\rho}$] the frequency response is quite broad and dips down sharply towards zero frequency for $D\gtrsim700$.  The discernible response feature indicated with an ellipse in Fig.~\ref{fig:Sqwtrapped}(f) is due to the roton mode  identified in Fig.~\ref{fig:Sqtrapped}(d) (but also has contributions from similar modes with $|m_z|>0$).

For the case of BECs with contact interactions a successful analytic approximation for  $S(k,\omega)$  has been developed using the Thomas-Fermi  approximation for the condensate and treating  the excitations within LDA  \cite{Zambelli2000a}.  We have found that the main   issue in extending this type of analysis to the tightly confined dipolar BEC arises from the sensitive dependence of the in-plane DDI potential [c.f.~Eq.~(\ref{V2D})]  upon the shape of the condensate in the  $z$ direction. 
For this reason we use the GPE solution itself as the basis  for calculating $S(k,\omega)$ using an LDA treatment of the excitations, thus avoiding the need to numerically solve for the BdG equations. We note that generalized $z$ mode treatments, e.g.~Ref.~\cite{Wilson2011a}, could also be used. We define a locally varying in-plane interaction potential
\begin{equation}
\!\tilde{V}_{\mathrm{2D}}^\prime({{k}_\rho},\rho)\!= \!\int\! \frac{dk_z}{2\pi n^2(\rho)}\!\left[g +\tilde{U}_{\mathbf{dd}}(k_{\rho},k_z)\right][\tilde{n} (k_z,\rho)]^2,\label{V2DGPE}
\end{equation} where $n(\rho)=\int dz\,|\psi_0(\rho,z)|^2$ is the areal density, $\tilde{n} (k_z,\rho)$ is the $z$-Fourier transform of the condensate density,  and $\tilde{U}_{\mathrm{dd}}(k_{\rho},k_z)=g_d(3k_z^2/k_{\rho}^2-1)$ is the Fourier transform of $U_{\mathrm{dd}}(\mathbf{r})$. The $\rho$ dependence of $\tilde{V}_{\mathrm{2D}}^\prime({{k}_\rho},\rho)$ accounts for the changing $z$ profile of the condensate as $\rho$ varies.
Note that in the limit of vanishing interactions, where the $z$ shape of the condensate is a Gaussian  (independent of $\rho$), $\tilde{V}_{\mathrm{2D}}^\prime({{k}_\rho},\rho)$ reduces to the analytic result in Eq.~(\ref{V2D}).
We construct $S(k,\omega)$ treating the in-plane excitations with the LDA, i.e.~summing over the parts of the BEC at various densities  \cite{Zambelli2000a}
\begin{equation}
S_{\mathrm{LDA}}(k,\omega)\!=\!\int d\rho\,2\pi\rho\,\frac{n(\rho)\epsilon(k_{\rho})}{\epsilon_{\mathrm{B}}( {k}_\rho,\rho)}
\delta\left(\omega\!-\!\epsilon_{\mathrm{B}}( {k}_\rho,\rho)/\hbar\right),\label{Sqw_GPELDA}\end{equation}
where $\epsilon_{\mathrm{B}}( {k}_\rho,\rho)=\sqrt{\epsilon(k_{\rho})^2 +2\epsilon(k_{\rho}) {n}(\rho)\tilde{V}^\prime_{\mathrm{2D}}({{k}_\rho},\rho) }$. 
In Figs.~\ref{fig:Sqwtrapped} (f1) and (f2) we compare our GPE based LDA (\ref{Sqw_GPELDA}) against   LDA calculation using the quasi-2D approximation [this only differs by the replacement $\tilde{V}^\prime_{\mathrm{2D}}({{k}_\rho},\rho) \to\tilde{V}_{\mathrm{2D}}({{k}_\rho})$ in Eq.~(\ref{Sqw_GPELDA})]. This comparison reveals the sensitivity of $S(k,\omega)$ to the $z$ shape of the condensate. In the inset to Fig.~\ref{fig:Sqtrapped}(e) we compare the full numerical calculations of $S(k)$ against the GPE based LDA, and find good agreements until $D\gtrsim 700$ where the roton modes approach zero energy.

Finally, we relate the dimensionless parameters of our calculations to current experimental systems. A value of $D\sim700$ for $\omega_\rho=2\pi\times40$ s$^{-1}$ would require a condensate with  $N=\{64.3,4.1,8.1\}\times10^4$ atoms for $\left\{^{52}\mathrm{Cr},\,^{164}\mathrm{Dy},\,^{168}\mathrm{Er}\right\}$, respectively.
The value of $D$ can be adjusted by changing the radial confinement, atom number or dipolar strength \cite{Giovanazzi2002a}. Our results demonstrate [see Fig.~\ref{fig:contours}(a)] that instead, for fixed dipole strength, the roton spectrum can be accessed by making the $s$-wave scattering length negative (e.g.~see \cite{Koch2008a}).

 In conclusion we have explored the excitation properties of a quasi-2D dipolar BEC in terms of the dynamic and static structure factors. Our results show that clear and direct signatures for the roton spectrum will emerge in the structure factors and should be readily observable with Bragg spectroscopy in current experiments. We have constructed  an approximate LDA theory for $S(k,\omega)$ which we have validated against the full theory. Future work will consider the extension to  a set of quasi-2D traps realized with an optical lattice \cite{Klawunn2009a,*Muller2011a,Wilson2011a}.

We acknowledge support from the Marsden Fund of New Zealand  contract UOO0924 and MSI contract  UOOX0915.

%

\end{document}